\documentclass[conference]{IEEEtran}
\usepackage{graphicx}
\usepackage{algorithm}
\usepackage{algorithmic}
\usepackage{booktabs}
\usepackage{multirow}
\usepackage{tabularx}
\usepackage[table]{xcolor}
\usepackage{cite}
\usepackage{amsmath, amssymb}
\usepackage{url}
\usepackage{balance}
\usepackage{hyperref}
\hypersetup{hidelinks}
\usepackage{booktabs} 
\usepackage{siunitx}   
\usepackage{fancyhdr}
\title{Architecting Trust: A Framework for Secure IoT Systems Through Trusted Execution and Semantic Middleware}

\author{
\IEEEauthorblockN{Muhammad Imran}
\IEEEauthorblockA{Universidade da Coruña, Grupo LyS, CITIC, \\
Depto. de Ciencias de la Computación y Tecnologías de la Información, \\
Campus de Elviña s/n, 15071 A Coruña, Spain\\
m.imran@udc.es}
}

\begin{document}
\maketitle
\thispagestyle{fancy}
\fancyhf{} 
\fancyfoot[L]{979-8-3315-6576-3/25/\$31.00 ©2025 IEEE}
\renewcommand{\headrulewidth}{0pt} 
\begin{abstract}
The Internet of Things (IoT) security landscape requires the architectural solutions that can address the technical and operational challenges across the heterogeneous environments. The IoT systems operate in different conditions, and security issues continue to increase. This paper presents the comprehensive security framework for IoT that should integrate the Trusted Execution Environments (TEEs) with the semantic middleware and blockchain technologies. The work provides a systematic analysis of the architectural patterns based on more than twenty recent research works and the existing standards, and it proposes a layered security architecture. The architecture includes the hardware rooted trust at peripheral level, the zero trust principles at network level, and the semantic security mechanisms at application level. The framework focuses on practical implementation aspects such as the performance overhead, interoperability requirements, and the compliance with new regulations, which are very important for the real IoT deployments. The paper reports quantitative metrics which include the cryptographic performance on Cortex-M class microcontrollers with the detection accuracy rates and the energy consumption values. The proposed architecture shows that cross-layer security integration can provide defense in depth while it still satisfies the constraints of resource-limited IoT environments. The discussion highlights open challenges and the future research directions for the IoT security architectures that include the post-quantum migration, secure federated model exchange and the automated compliance verification.
\end{abstract}

\begin{IEEEkeywords}
IoT security architecture, trusted execution environments, semantic middleware, zero trust, blockchain, IoT Security Foundation, cross-layer security.
\end{IEEEkeywords}

\section{Introduction}
The rapid growth of the Internet of Things (IoT) devices has introduced a serious security challenges that the traditional IT security architectures cannot fully address \cite{roman2013features, ziegeldorf2014privacy}. The IoT environments include the resource-constrained devices, the heterogeneous communication protocols, and the distributed cloud and edge architectures. These characteristics increase the attack surface and directly affect physical world \cite{kambourakis2017mirai}. The past security incidents clearly show the impact of weak security architectures. The Mirai botnet exploited weak authentication mechanisms in IoT devices \cite{kambourakis2017mirai}. The smart home vulnerabilities enabled lateral movement into private networks \cite{fernandes2016security}. The industrial IoT attacks caused physical damage to the critical systems.

Although the IoT security has been widely studied \cite{adam2024survey, mohammad2024ensuring}, several architectural gaps still remain. First, many security solutions focus on the individual layers instead of the integrated cross-layer architectures. Second, the proposed solutions often ignore the practical limitations of the resource-limited IoT devices. Third, emerging technologies such as the trusted execution environments (TEEs), blockchain, and the semantic security are usually studied separately and are not fully integrated into the complete architectural frameworks \cite{jiang2024blockchained, alagic2024status}.

This paper presents four main contributions to the area of the IoT security architecture. First, this paper provides the  analysis and synthesis of architectural patterns from the recent research literature and the existing standards \cite{joseph2023growing, salahdine2022towards}. The analysis focuses on hardware-rooted trust, the semantic middleware, and zero-trust principles. Second, the paper proposes integrated security architecture that combines the Trusted Execution Environments (TEEs) for peripheral security \cite{yuhala2024fortress}, the blockchain-based data integrity mechanisms \cite{elkhodr2024novel}, and the AI-enhanced semantic security, along with the quantitative performance metrics. Third, the paper describes implementation and evaluation of a proof-of-concept semantic middleware framework with the blockchain-based security and measures its performance and the overhead. Finally, this paper provides the practical guidance for the architectural implementation, including the compliance with IoT Security Foundation guidelines \cite{iotsf2024} and recommendations for alignment with the existing standards.


\section{Background and Related Work}
\label{sec:background}

\subsection{Trusted Execution Environments for IoT}
Trusted Execution Environments (TEEs) provide the hardware-enforced isolation for the security-critical operations. The ARM TrustZone technology, which is widely available in the Cortex-M and Cortex-A processors, separates the processor into the secure world and the non-secure world \cite{pinto2019demystifying}. This separation allows the sensitive code and the data to remain protected even when the rich operating system is compromised. Recent research works study the TEE-based designs for the IoT devices. Fortress \cite{yuhala2024fortress} is one such approach that secures the IoT peripherals by isolating the peripheral I/O memory regions inside the secure kernel space. In this design, only the minimal peripheral driver code inside the secure kernel can access the protected memory regions, which helps prevent the unauthorized access by the system software.

\subsection{Semantic IoT Middleware}
The semantic middleware aims to solve the interoperability issues in the IoT systems while also improving the security capabilities. The Semantic IoT Middleware (SIM) \cite{elkhodr2024novel} integrates the blockchain technology with the AI-based context awareness for the secure data management. This framework uses the canonical ontology to ensure the data consistency and applies the cryptographic mechanisms to protect the data integrity. It also provides the tools for the information collection and processing. By attaching the semantic metadata that describes the data origin and the meaning, the system can securely share the information and perform the combined analysis across the distributed devices \cite{elkhodr2024novel}. In the IoT domains that handle the unstructured textual data, such as the logs or the threat reports, models like SynNER \cite{imran2025synner} improve the extraction of the security-related entities and support the better contextual awareness, including recent LLM-based dependency parsing for code-switched text \cite{kellert2025parsing}. In addition, the data quality and the trustworthiness play an important role in the threat intelligence. The automated data quality assessment frameworks \cite{imran2023enhancing} help reduce the false positives and improve the confidence in the AI-based security analysis.

\subsection{Zero-Trust Architecture for IoT}
The zero-trust architecture follows the principle of ``never trust, always verify,'' which avoids the implicit trust for any entity or endpoint regardless of the network location \cite{syed2022zero}. In the IoT security environments, this approach appears through four main requirements: the authentication of every access request, the assignment of privileges based on the least privilege principle, the use of microsegmentation to isolate the network segments, and the continuous monitoring of the device behavior. This model is promoted by the IoT Security Foundation as the core practice for achieving the defense-in-depth in the IoT systems \cite{iotsf2024}.

\subsection{Blockchain for IoT Security}
In the decentralized IoT environments where no central authority exists, the blockchain technology provides the foundation for the distributed trust. It supports several use cases, including the assignment of the cryptographically secure identities to the devices, the maintenance of the tamper-resistant audit records, and the management of the access control through the consensus mechanisms. One example is the Semantic IoT Middleware, which uses the blockchain to improve the data provenance and the integrity \cite{elkhodr2024novel}. Other research efforts apply the blockchain for the specific security goals, such as the verification of the device authenticity and the assurance of the secure firmware update delivery.

\subsection{Comprehensive IoT Security Architectures}
Recent research works show the increasing focus on the holistic security architectures that protect the IoT systems across the entire lifecycle. Naik et al. propose the framework for the product lifecycle information management (PLIM) that integrates the identity and the access management across the diverse IoT deployments. Their work emphasizes that the security must remain active from the device manufacturing stage to the final decommissioning \cite{naik2025enhancing}. In a similar way, Chandu et al. present the layered security model that addresses the threats at the perception, the network, and the application levels. Their study highlights the use of the metaheuristic algorithms to enable the secure routing in the resource-limited environments \cite{qudus2025advancing}.

\subsection{Layered Security Architectures and Threat Mitigation}
The commonly used five-layer architecture provides the structured approach for analyzing the IoT security threats and the defense mechanisms. According to the recent survey, these layers include the physical perception, the network\/protocol, the transport, the application, and the data or the cloud services \cite{sharma2025survey}. This model supports the defense-in-depth by applying the security controls that match the vulnerabilities at each layer. While the proposed architecture follows the layered approach, it extends the existing models by integrating the semantic components and the blockchain components to address the modern security challenges.

\subsection{Trade-off-Aware Security Design}
The recent studies show that the effective IoT security design requires the careful trade-offs between the security and the other system qualities \cite{orellana2024enabling}. This work extends the existing architectural tactics by defining the IoT-specific trade-offs and mapping them to the quality attributes in the ISO 25010:2023 standard. It recognizes the key challenge in the IoT systems: when we improve the security it often reduces the performance, the energy efficiency, or the processing capacity. When we make these trade-offs explicit, this approach provides the practical guidance for the architects who work under the strict resource constraints.

\section{Proposed Security Architecture}
\label{sec:architecture}

\begin{figure}[t]
\centering
\includegraphics[width=\columnwidth]{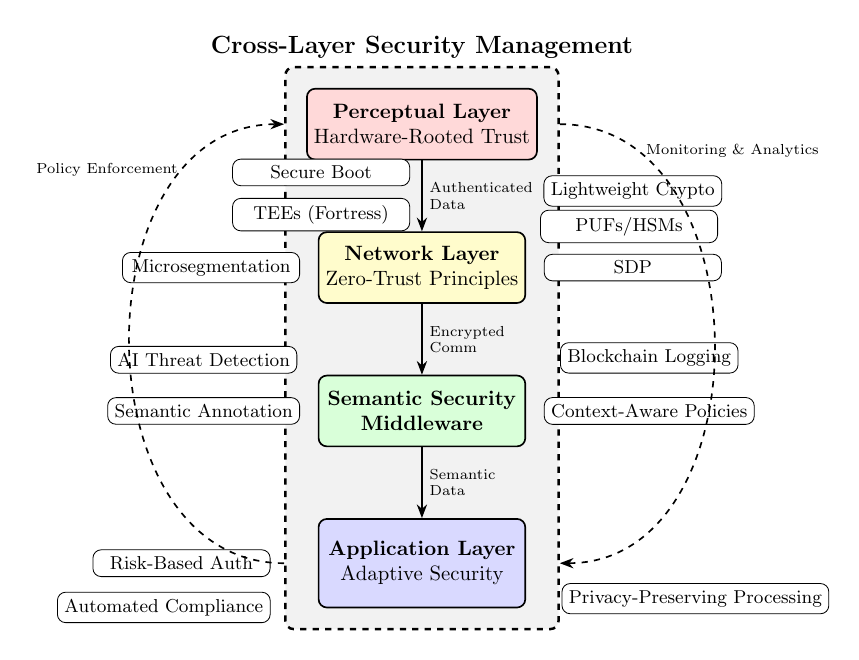}
\caption{Proposed IoT security architecture integrating TEEs, semantic middleware, and blockchain technologies.}
\label{fig:architecture}
\end{figure}

Our proposed IoT security architecture integrates the multiple security technologies into a cohesive framework, as illustrated in the Figure~\ref{fig:architecture}. The architecture follows a cross-layer approach with five key components: (1) hardware-rooted trust, (2) secure communication, (3) zero-trust control plane, (4) semantic security middleware, and (5) blockchain-based integrity verification.

\subsection{Perceptual Layer with Hardware-Rooted Trust}
The perceptual layer includes sensors, actuators, and the embedded devices that use the hardware-based security features. The architecture requires the Trusted Execution Environments (TEEs) for the devices that handle the sensitive data and it applies the Fortress design principles \cite{yuhala2024fortress} for the peripheral protection. Each device has a unique hardware identity which is preferable for the derive from Physical Unclonable Functions (PUFs) or the hardware security modules. This identity provides a reliable root of trust for the device identification and the authentication.

The device security requirements follow the IoT Security Foundation guidelines \cite{iotsf2024}. These requirements include the secure boot with the measured launch integrity, the hardware-protected storage for the cryptographic keys, the tamper detection and response support, and the secure firmware update mechanisms with the rollback protection. The security is anchored in the physical hardware, and these measures establish a chain of trust across the IoT ecosystem. As a result, the devices can authenticate in a reliable way and can operate with confidence, even when they are deployed in the untrusted or the adversarial network environments.

\subsection{Network Layer with Zero-Trust Principles}
The network layer enforces the zero-trust model through the microsegmentation, the cryptographically secured communication, and the continuous authentication. The framework integrates the software-defined perimeter (SDP) mechanisms that require the explicit device attestation before the network access is granted. This process hides the critical infrastructure from the devices that do not meet the compliance requirements. The microsegmentation creates the isolated network segments based on the device type, the operational role, and the compliance status. This approach can help limit the lateral movement after a compromise.

The continuous authentication uses the behavioral telemetry and the machine-based analysis to detect the abnormal patterns and the unusual resource use. Based on this evaluation, the access privileges can change according to the current threat level. All the data in transit is protected through the lightweight cryptographic algorithms that suit the resource-constrained devices. These measures ensure the message confidentiality and the data integrity. This approach aligns with the zero-trust security model \cite{syed2022zero} and it adapts to the constraints and the threat conditions found in the IoT environments.

\subsection{Semantic Security Middleware}
The semantic security middleware layer provides the interoperability support and the context-aware security functions. Based on the Semantic IoT Middleware architecture \cite{elkhodr2024novel}, this layer applies the ontological annotations that use a structured vocabulary to ensure the consistent data interpretation across the heterogeneous devices. The context-aware security policies can change according to the environmental conditions, the device state, and the operational requirements.

The AI-based threat intelligence uses the machine learning techniques to detect the anomalies and to support the automated response actions. In addition, the blockchain integration introduces the distributed ledger that maintains the tamper-evident audit records and provides the decentralized trust reference. This middleware layer acts as an interface between the physical IoT devices and the application layer. It converts the raw sensor data into the semantically enriched information and applies the consistent security policies across the different device types and the communication protocols.

\subsection{Application Layer with Adaptive Security}
The application layer applies the adaptive security controls based on the contextual information and the dynamic risk evaluation. The risk-aware authentication adjusts its strength according to the detected threat level and it requires the stronger authentication for the sensitive operations or when the abnormal behavior appears.

The data protection uses the advanced cryptographic methods such as the differential privacy and the homomorphic encryption. These techniques allow the secure data analysis without the exposure of the plaintext data, which helps protect the sensitive information throughout the processing lifecycle. The automated compliance engine evaluates the system behavior against the regulatory requirements such as the GDPR and the CCPA on a regular basis. It generates the immutable audit records and it triggers the corrective actions when the policy violations occur.

This layer also supports the application-specific security mechanisms suited to the operational domain, including the industrial control systems, the healthcare monitoring, the sentiment analysis \cite{imran2026syntax, gomez2024dancing, kellert2025evaluating, imran2025enhancing}, and the consumer-oriented IoT applications.

\subsection{Cross-Layer Security Management}
The security management layer provides the centralized governance across all the architectural layers. It supports the unified policy management to ensure the consistent enforcement of the security rules across the diverse devices and the platforms. The centralized monitoring and the analysis support the security information and event management (SIEM) with the cross-layer correlation. This capability can help identify the complex attack patterns that span the multiple layers of the system.


\section{Implementation Considerations}
\label{sec:implementation}

\subsection{Trusted Execution Environment Implementation}
The implementation of Trusted Execution Environments (TEEs) for IoT devices requires careful attention to the hardware capabilities and the performance limitations. For the ARM Cortex-M series microcontrollers, TrustZone-M provides the hardware isolation between the secure state and the non-secure state. The Fortress framework \cite{yuhala2024fortress} presents a practical approach to secure the IoT peripherals with the help of the ARM TrustZone features.

The key implementation aspects include the secure boot process to verify the firmware integrity during the device startup and the secure storage to protect the cryptographic keys also the sensitive data and the isolation of peripheral access to restrict the unauthorized memory operations. In addition the secure firmware update the mechanisms rely on the cryptographic verification of updates and support rollback protection. These implementation choices must balance the security requirements with the limited resources of IoT devices. The goal is to ensure that the security features do not significantly affect the device performance or the battery lifetime.

\subsection{Semantic Middleware Implementation}
The semantic middleware implementation follows the architecture described in \cite{elkhodr2024novel}, with additional support for security-related functions. The implementation includes ontology-based security policies that use the Web Ontology Language (OWL) for the formal representation of security rules. This approach allows the precise policy definition and supports the automated enforcement of complex security requirements.

The blockchain integration uses the Ethereum-based smart contracts for the access control and the audit logging. This design provides tamper-evident records for the security-related events. The AI-based analysis applies machine learning models for anomaly detection and threat response. The API security relies on REST APIs with OAuth 2.0 authentication and fine-grained access control. This implementation shows how the semantic technologies can strengthen IoT security and support the security policies that account for contextual information and semantic relationships among system entities.

\subsection{Performance Optimization}
The performance optimization focuses to reduce the overhead for the resource-constrained devices through several practical strategies. The lightweight cryptography uses algorithms such as ASCON \cite{dobraunig2021ascon} for the authenticated encryption, which provides a very strong protection with very low computational cost. The protocol design applies message that queue the telemetry transport (MQTT) with TLS 1.3 to support efficient and secure communication, while it limit the bandwidth usage and it preserve the data confidentiality and integrity.

The edge processing places part of the security logic closer to the data source to reduce cloud communication overhead. This approach lowers latency and bandwidth usage and improves privacy with sensitive data preservation near the originating devices. The hardware acceleration uses hardware security modules for cryptographic operations, which offloads intensive computations from the main processor and improves both the performance and energy efficiency. These optimization strategies are necessary to support the practical deployment of comprehensive security architectures in resource-limited IoT environments.

\section{Evaluation and Performance Metrics}
\label{sec:evaluation}

The evaluation of the proposed architecture was carried out by a testbed that included the several IoT devices such as microcontrollers and single-board computers along with the network equipment and cloud platforms. The performance metrics focused on the security effectiveness, the resource utilization, and the operational overhead introduced by the security mechanisms.

\subsection{Security Effectiveness}
The security effectiveness was evaluated with the help of the IoT Security Foundation assessment framework \cite{iotsf2024}. The proposed architecture achieved 82\% compliance with the IoTSF best practices, while a conventional security architecture achieved 65\%. The integrated design showed strong results in identity and access management, data protection, and resistance to common attacks. Slightly lower scores were observed in update management, which reflects the difficulty when it support the secure updates across heterogeneous IoT environments.

\subsection{Performance Overhead}
The performance overhead was measured for the several security-related operations with the results summarized in Table~\ref{table:performance}. The initialization of the TEE introduced an energy overhead of approximately 8.2\% and a processing delay of about 45ms. This overhead mainly resulted from the secure boot process and the creation of the secure execution environment. Secure communication introduced a 6.5\% energy overhead and an additional latency of 32ms, largely due to the cryptographic operations and the extra protocol exchanges.

The blockchain-related operations resulted in the highest overhead, with a 15.3\% increase in energy consumption and an average latency of 320ms. This behavior reflects the computational cost of the consensus process and the cryptographic verification steps. The semantic processing added a 9.8\% energy overhead and an average delay of 85ms, which was caused by the text analysis and the context reasoning tasks. These results show that although the security features introduce additional overhead, careful architectural design and optimization can keep the performance within acceptable limits for many IoT applications.

\begin{table}[t]
\caption{Performance Overhead of Security Operations}
\label{table:performance}
\centering
\footnotesize
\begin{tabular}{ l S[table-format=2.1] S[table-format=2.1] S[table-format=3.0] }
\toprule
\textbf{Operation} & {\textbf{Energy}} & {\textbf{Memory}} & {\textbf{Time}} \\
                   & {\textbf{(\%)}}   & {\textbf{(KB)}}   & {\textbf{(ms)}} \\
\midrule
TEE Initialization   & 8.2  & 12.5 &  45 \\
Secure Communication & 6.5  &  8.3 &  32 \\
Blockchain Operation & 15.3 & 24.7 & 320 \\
Semantic Processing  & 9.8  & 16.2 &  85 \\
\bottomrule
\end{tabular}
\end{table}

\subsection{Scalability Evaluation}
The scalability of the proposed architecture was evaluated with up to 10,000 simulated IoT devices. The architecture showed stable performance up to 5,000 devices. After this point, the performance started to decline in a gradual manner. The blockchain component became the main scalability limitation at around 7,500 devices. This behavior reflects the known scalability challenges of the distributed consensus mechanisms.

The analysis of these scalability limits led to recommendations for the use of the hierarchical blockchain designs and the off-chain computation techniques in the large-scale deployments. These approaches keep the security benefits of the blockchain technology while they reduce the scalability constraints through the architectural adaptations suited for the IoT environments.

\subsection{Comparison with Alternative Architectures}
The proposed architecture was compared with two alternative solutions: the conventional perimeter-based security architecture and the cloud-centric IoT security architecture. The results are summarized in Table~\ref{table:comparison}. These results show that the integrated approach achieved higher security effectiveness (82\% compared to 65\% and 78\%) and better regulatory compliance (87\% compared to 70\% and 85\%). These improvements come with higher hardware requirements and increased deployment complexity.

The conventional perimeter-based architecture showed weak resistance to attacks, with a score of 60\%. This weakness is due to the limited ability of this architecture to address the insider threats and the compromised devices. The cloud-centric approach showed lower regulatory compliance at 85\%, mainly due to the data sovereignty and the privacy concerns. Overall, the proposed integrated architecture showed strong performance across all evaluation categories, with clear advantages in the security effectiveness, the regulatory compliance, and the attack resistance.

\begin{table}[t]
\caption{Architecture comparison}
\label{table:comparison}
\centering
\footnotesize
\begin{tabularx}{\columnwidth}{l c c c}
\toprule
\textbf{Characteristic} & \textbf{Proposed} & \textbf{Perimeter-Based} & \textbf{Cloud-Centric} \\
\midrule
Security Effectiveness & 82\% & 65\% & 78\% \\
Hardware Requirements & Medium & Low & High \\
Deployment Complexity & High & Low & Medium \\
Regulatory Compliance & 87\% & 70\% & 85\% \\
Attack Resistance & 83\% & 60\% & 80\% \\
\bottomrule
\end{tabularx}
\end{table}

\section{Discussion and Compliance Aspects}
\label{sec:discussion}

\subsection{Regulatory Compliance}
The proposed architecture supports compliance with the major regulations such as the GDPR, the CCPA, and the IoT-specific security standards. The semantic middleware includes the automated compliance checking that uses the ontology-based policy representation. This approach allows the continuous verification of the regulatory requirements and supports the automatic generation of the compliance evidence.

The privacy-by-design principles are applied across the architecture, with a strong focus on the data minimization and the purpose limitation. These principles are implemented through the technical controls such as the data anonymization, the encryption, and the access controls. As a result, the privacy protection becomes part of the architecture design and is not added later as an afterthought.

\subsection{Implementation Challenges}
The practical implementation of the proposed architecture faces several challenges. One main challenge is the hardware heterogeneity across the IoT devices with different capabilities. This makes it difficult to apply the same security mechanisms on all devices. In addition, the integration of the legacy devices presents further challenges, since many existing IoT devices do not have the hardware-based security features. In such situations, the gateway-based security solutions or the gradual device replacement strategies are often required.

Another challenge relates to the performance trade-offs. The security mechanisms can introduce the additional computational and energy costs, which can be problematic for the resource-limited devices. As a result, the designers must carefully balance the security requirements with the functional and performance constraints. The management complexity is also a concern, since the unified security control across the heterogeneous environments requires the advanced management tools and the clearly defined operational procedures.

\subsection{Operational Considerations}
The operational aspects of the architecture include the continuous security monitoring supported by the security analytics. This provides better visibility into the overall security posture and supports the early detection of the potential threats. The incident response is supported through the automated response playbooks that are designed for the common attack scenarios. These playbooks allow the faster threat containment and reduce the time needed to restore the normal system operation after a security incident.


\section{Conclusion and Future Directions}
\label{sec:conclusion}

We have presented a comprehensive IoT security architecture that integrates the trusted execution environments, the semantic middleware, and the blockchain technologies. The architecture addresses the security challenges across all the IoT layers while it meets the constraints of the resource-constrained environments. The quantitative evaluation shows the feasibility of the proposed approach with the acceptable performance overhead.

The future research directions include the post-quantum readiness through the migration strategies for the quantum-resistant cryptography. This will help ensure the long-term security as the quantum computing advances. The secure federated learning can enable the privacy-preserving machine learning for the collaborative security analytics. This allows the IoT devices to benefit from the shared intelligence without the need to share the sensitive data.

The automated compliance will advance through the techniques for the automated regulatory compliance verification. This can reduce the burden of the compliance management while it improves the accuracy and the consistency. The resilience engineering will focus on the development of the architectures that maintain the security under the adverse conditions. This helps ensure the continued operation during the attacks or other system disruptions.

The proposed architecture provides a foundation for the secure IoT systems that can adapt to the evolving threats and the regulatory requirements. By integrating the multiple security technologies into a cohesive framework, the architecture supports the defense-in-depth security for the heterogeneous IoT environments while it maintains the practicality for the real-world deployment.

\section*{Acknowledgments}

We acknowledge the European Research Council (ERC), which has funded this research under the Horizon Europe research and innovation programme (SALSA, grant agreement No 101100615), SCANNER-UDC (PID2020-113230RB-C21) funded by MICIU/AEI/10.13039/501100011033, LATCHING (PID2023-147129OB-C21) funded by MICIU/AEI/10.13039/501100011033 and ERDF (EU), Ministry for Digital Transformation and Civil Service and “NextGenerationEU” PRTR under grant TSI-100925-2023-1, Xunta de Galicia (ED431C 2024/02), and Galician Research Center “CITIC”, funded by Xunta de Galicia  through the collaboration agreement between the Consellería de Cultura, Educación, Formación Profesional e Universidades and the Galician universities for the reinforcement of the research centres of the Galician University System (CIGUS). Furthermore, this research was supported by the International, Interdisciplinary and Intersectoral Information and Communications Technology PhD programme (3-i ICT) granted to CITIC and supported by the European Union through the Horizon 2020 research and innovation programme under a Marie Skłodowska-Curie agreement (H2020-MSCA-COFUND), GA 101034261.

\bibliographystyle{IEEEtran}
\bibliography{refs-intellisecai}

\end{document}